\documentclass[aps,prb,superscriptaddress,twocolumn,amsmath,amssymb,floatfix]{revtex4-1}
\pdfoutput=1
\usepackage{bm}
\usepackage{graphicx}
\usepackage{amsfonts}
\usepackage{amsbsy}

\begin{document}
 
\title{Higher angular momentum pairing from transverse gauge interactions}
\author{Suk Bum Chung$^1$, Ipsita Mandal$^1$, Srinivas Raghu$^2$, and Sudip Chakravarty}
\affiliation{Department of Physics and Astronomy, University of California Los Angeles, Los Angeles, California 90095, USA\\
$^2$Department of Physics, Stanford University, Stanford, California 94305, USA}
 
\date{\today}
 
\begin{abstract}
In this paper, we study superconductivity of nonrelativistic fermions at finite-density coupled to a transverse $U(1)$ gauge field, with the effective interaction including the Landau-damping. This model, first studied by Holstein, Norton, and Pincus [Phys. Rev  B, {\bf 8}, 2649 (1973)]  has been known as an example of a non-Fermi liquid, {\i.e.} a metallic state in which the decay rate of a quasiparticle is large compared to the characteristic quasiparticle energy; other examples of the non-Fermi liquid includes the 2d electron gas in a magnetic field at $\nu=1/2$ and the normal state of optimally doped cuprate superconductors. Our study thus addresses the question of whether or not non-Fermi liquids, like Fermi liquids, are unstable towards the formation of superconductivity.
The results are (i) the non-Fermi liquid is stable against superconductivity below a critical gauge coupling, (ii) above this critical coupling, the ground state is an unconventional superconductor with angular momentum $\ell \ge 2$. Our results are obtained from a solution of the Dyson-Nambu equation. Note that in this problem there is a quantum critical point between a non-Fermi liquid state and the superconducting state, as the critical coupling is nonzero. This is in contrast to a weakly coupled metal, which exhibits superconductivity for infinitesimally weak interaction regardless of its sign.

\end{abstract}
 
\maketitle
 
\section{Introduction}
A classic problem of condensed matter physics is that of a finite density of electrons coupled to transverse gauge fields. As was first noticed by Holstein, Norton, and Pincus (HNP) \cite{Holstein:1973} and elaborated further by Reizer \cite{Reizer:1989}, such a system exhibits anomalous properties, since the transverse component of the gauge field remains unscreened. While the original motivation of HNP was to understand the effects of the electromagnetic field coupled to a metal, the same field theory applies to the case of fermions coupled to an emergent gauge field, as is the case in certain spin-liquids, and perhaps even the normal state of cuprate superconductors \cite{Baskaran:1988, Ioffe:1989, Lee:1989, Blok:1993, Ubbens:1994,Lee:1998} and compressible quantum Hall systems at the filling fraction of $1/2$.~ \cite{Halperin:1993, Nayak:1994} As is true for any metal, the system possesses infinitely many gapless excitations; however the decay rate of these ``quasiparticles" is parametrically larger then their characteristic energy. A largely unsolved issue is whether or not such a non-Fermi liquid metal can remain stable against the formation of superconductivity.
 
While the treatments of HNP \cite{Holstein:1973}, and Reizer \cite{Reizer:1989}, were based on the random phase approximation (RPA), an analysis based on renormalization group~\cite{Chakravarty:1995} treatment in the vicinity of the upper-critical dimension, $\epsilon =(3-d)$, suggested that the problem of fermions coupled to transverse gauge bosons leads to a non-Fermi liquid fixed point, also found by Nayak and Wilczek \cite{Nayak:1994}, who considered a more general expansion in $\epsilon =3 -(d+x)$, where $x$ is the exponent characterizing the range of the four fermion interaction. The stability of the non-Fermi liquid fixed point with respect to competing ordered states, however, remains an outstanding issue. The issue has also  been addressed in the context of color superconductivity~ \cite{Alford:2008}; in the simplest case up and down quarks with two different flavors form $s$-wave superconductivity whereas the quarks of the third color remain gapless.
 
Here we consider $T=0$ and show that emergent transverse $U(1)$ gauge bosons, more appropriate to condensed matter physics,   coupled to fermions in a fermi liquid can open up superconducting gap. Once a gap is present, all non-Fermi liquid
anomalies are cut off and the theory is self consistent. The quantum phase transition is a continuous transition at a nonzero critical gauge coupling. This is unlike conventional theory of superconductivity where arbitrarily small attractive coupling leads to superconductivity. It is also unlike Kohn-Luttinger superconductivity \cite{Kohn:1960} where infinitesimally small repulsive coupling leads to superconductivity.

Our key results here are twofold. Firstly, the non-Fermi liquid metal resulting from coupling a Fermi surface to a transverse $U(1)$ gauge field remains a stable ground state below a critical threshold gauge coupling; there is no superconducting instability. Above the threshold coupling, there are  instabilities towards  unconventional superconducting states with angular momentum $\ell \ge 2$. Our analysis is carried out via a straightforward Hatree-Fock solution of the self-consistent gap equation. While it neglects fluctuation effects, it provides a simple qualitative guideline to the possible nature of the ground states, and is likely to remain reliable close to $d=3+1$, which is the upper-critical dimension for the problem at hand; the problem we consider corresponds to $x=0$. We also briefly touch upon the case $d=2+1$.

This paper is organized as follows. In Sec. II, we present our system of nonrelativistic fermions at finite-density coupled to a transverse $U(1)$ gauge field. In Sec. III, we derive the Hartree-Fock superconducting gap equation from the Dyson formalism. In Sec. IV we present our numerical solution to the gap equation. Lastly we discuss the implications and the validity of our results. In an Appendix we  consider a similar treatment at $d=2+1$, which we believe is still above the lower critical dimension, and our Hartree-Fock treatment could still provide a qualitative guide.
 
\section{The Model system}
 
We first present our action for the fermions coupled to the transverse gauge bosons at $d=3$. Our action is Wick-rotated at $T=0$, 
which gives us the form most convenient for investigating the quantum phase transition from a non-Fermi liquid to a superconducting state. The full action consists of three main terms - the free fermion action denoted by $\psi$'s, the free $U(1)$ gauge field, denoted by $A$'s, and 
the fermion-boson coupling ($i,j=1,2,3$):
\begin{widetext}
\begin{eqnarray}
S_{F}&=&\int \frac{d\omega}{2 \pi} \int \frac{d^3 k}{(2 \pi)^3} \psi^{\dagger}({\bf k},\omega)\left[i\omega- \left(\varepsilon_k -\mu\right)\right]\psi({\bf k},\omega),\\
S_{G}&=& \int \frac{d\omega}{2 \pi} \int \frac{d^3 q}{(2 \pi)^3}A^{\dagger}_{i}({\bf q},\nu)\left[q^{2}+\nu^{2}\right]\left(\delta_{ij}-q_{i}q_{j}/q^{2}\right)A_{j}({\bf q},\nu),\\
S_{\text{int}} &=&\frac{g}{m} \int \frac{d\omega}{2 \pi} \int\frac{d\nu}{2 \pi} \int \frac{d^3 p}{(2 \pi)^3} \int \frac{d^3 k}{(2 \pi)^3} \psi^{\dagger}({\bf k},\nu) \lbrace A_{i}({\bf k}-{\bf p}, \nu-\omega)p_{i} \rbrace \, \psi({\bf p},\omega),
\end{eqnarray}
\end{widetext}
where repeated indices are summed over, $m$ is the mass of the fermion, $\varepsilon_k = k^2/2m$ and $g$ is the gauge coupling. We work in the Coulomb gauge such that $\partial_{i}A_{i}=0$. To simplify notation we have set $\hbar$ and the gauge boson velocity $c_{g}$ to unity. We have subsumed all four fermion interactions into $S_F$ in the spirit of Landau theory. The gauge group is assumed to be noncompact, which is not entirely a trivial assumption. In $(2+1)$-dimensions, Polyakov~\cite{Polyakov:1987} has shown that the gauge field is massive for the compact case due to instanton effects, at least in the absence of the matter field. In the presence of massless fermions and in $(3+1)$ dimensions, such topological excitations are assumed not to play any important role.~\cite{Khlebnikov:1994}
 
\begin{figure}[h]
\centering
\includegraphics[width=\linewidth]{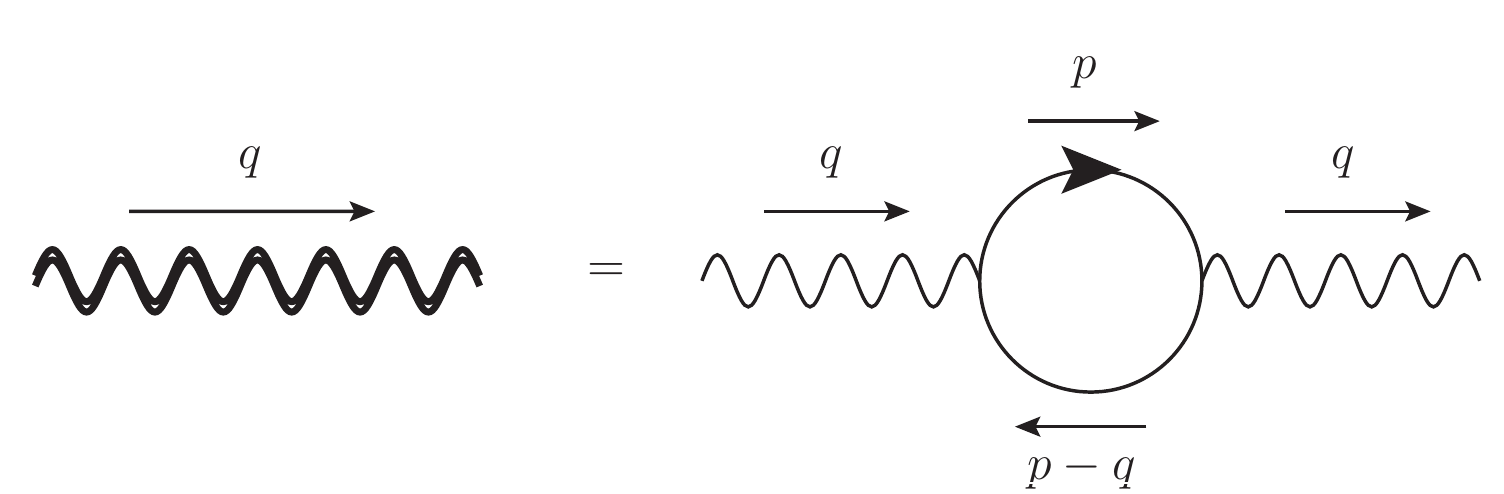}
\caption{One-loop corrected gauge boson propagator. Here for compactness we use Euclidean four vector notation: $p=(p_{0},{\bf p})\equiv (\omega,{\bf p})$.}
\label{fig:boson}
\end{figure}

Since we are investigating the possibility of superconductivity cutting off the non-Fermi liquid behavior, we use for our effective transverse gauge boson mediated interaction the same effective interaction that is known to give rise to the non-Fermi liquid behavior in this system \cite{Chakravarty:1995}. 
It was shown previously that this effective interaction can be obtained simply by taking the one-loop corrected gauge boson propagator as 
the vertex does not receive any anomalous dimension to one-loop order from the Ward identity~\cite{Chakravarty:1995} a conclusion that was also reached by
Polchinski.~\cite{Polchinski:1994}
In other words, our problem is to find out whether the interaction mediated by the gauge bosons with the one-loop corrected effective action
\begin{widetext}
\begin{equation}
S_{G}= \int \frac{d\nu}{2 \pi} \int \frac{d^3 q}{(2 \pi)^3}A^{\dagger}_{i}({\bf q},\nu)\left[q^{2}+\nu^{2}+\gamma |\nu|/qv_{F}\right]\left(\delta_{ij}-q_{i}q_{j}/q^{2}\right)A_{j}({\bf q},\nu),
\end{equation}
\end{widetext}
can 
lead to superconductivity in $(3+1)$ dimensions. The corresponding gauge propagator 
\begin{equation}
D_{ij} ({\bf k}, \nu) = \big ( \delta_{ij} - \frac{k_i k_j}{k^2}\big) \frac{1}{k^2 + \gamma \frac{|\nu|}{k v_{F}}}
\end{equation}
is shown graphically in Fig.~\ref{fig:boson}.
where $v_{F}$ is the Fermi velocity, $\gamma = \alpha k_{F}^{2}$, and $\alpha = \frac{g^{2}v_{F}}{4\pi}$ is our analog of the fine structure constant; 
note that we have dropped the $\nu^{2}$ term, keeping only the most relevant low frequency term that arises as the result of the Landau damping. It is important to note that our fine structure constant $\alpha$ can be much larger than that of quantum electrodynamics, given that $v_F$ can be much larger than the gauge velocity, set typically by a strongly correlated many body system, and of course $g$ itself is not known in general.
 
Our problem is analogous to the Kohn-Luttinger problem, albeit with some qualifications. To see how this is so, note that the transverse gauge bosons mediate current-current interaction. The bare current-current interaction in the Cooper channel is repulsive for the same reason as two parallel wires with anti-parallel currents repel each other. Thus, our problem shares the most essential characteristic of the Kohn-Luttinger problem - the Cooper pairing out of a repulsive bare interaction. However, whereas in the Kohn-Luttinger problem the infinitesimal bare interaction is sufficient to destabilize the Fermi liquid through Cooper pairing, in our problem a finite interaction strength - that is to say, a finite coupling constant - is required for Cooper pairing to destabilize the non-Fermi liquid \cite{Schafer:2006}. This is because the screening of our interaction is dynamical, meaning that we recover the bare interaction in the zero frequency limit. Together with the fact that at the zero frequency limit we will have equally divergent repulsion in all angular momentum channels, 
this makes the perturbative method used in the Kohn-Luttinger problem quite unsuitable for our problem.
 
\section{Gap equation from Dyson formalism}

\begin{figure}[h]
\centering
\includegraphics[width=\linewidth]{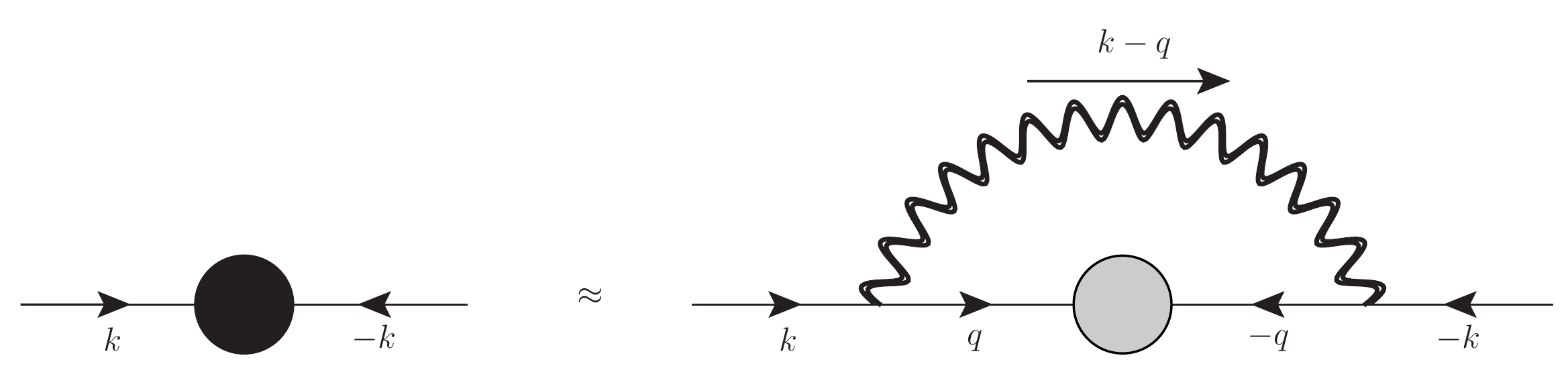}
\caption{Single-particle-exchange contribution to the off-diagonal component of the Dyson equation. The black blob denotes the proper self-energy, and the grey blob (along with the associated fermion lines) denotes the exact propagator of the fermions. As in Fig.\ref{fig:boson}, we have used the Euclidean four vector notation for simplicity.}
\label{fig:dyson}
\end{figure}

We obtain the pairing gap self-consistently through Dyson formalism, as this technique can be applied straightforwardly to a frequency-dependent interaction \cite{Eliashberg:1958, Nambu:1960, Culler:1962, Schrieffer:1963, Schrieffer:1983}. In this formalism, we treat the pairing gap as the self-energy of the fermion propagator in the Nambu basis, as seen in Fig.~\ref{fig:dyson}.
In our Dyson equation,
\begin{align}
\Sigma({\bf k},\omega) =\int \frac{d^3 q d\nu}{(2\pi)^4} V_{\text{eff}}({\bf k},\omega; {\bf q},\nu) G({\bf q},\nu),
\end{align}
the self-energy $\Sigma({\bf k},\omega)$ is due to the gauge boson mediated interaction in the Cooper channel
\begin{align}
V_{\text{eff}}({\bf k},\omega; {\bf q},\nu) =&- \frac{g^2}{m^2} (k_i) D_{ij} ({\bf k}-{\bf q}, \omega-\nu) (-k_j)\nonumber\\
=& \frac{g^2}{m^2} \frac{k^2-\frac{[{\bf k}\cdot({\bf k}-{\bf q})]^2}{({\bf k}-{\bf q})^2}}{({\bf k}-{\bf q})^2 + \gamma \, \frac{|\omega-\nu|}{v_F |{\bf k}-{\bf q}|}}.
\label{EQ:effV1}
\end{align}
In order to derive the gap equation from this Dyson equation, we note that the self-energy by definition is the difference between the free and interacting propagator:
\begin{equation}
-\Sigma({\bf k},\omega) =G^{-1}({\bf k},\omega)-G^{-1}_0({\bf k},\omega).
\end{equation}
In the Nambu basis, the inverse non-interacting fermion propagator is
\begin{equation}
G^{-1}_0({\bf k},\omega) = \begin{pmatrix}
i \omega -\xi_k & 0 \\
0 & i \omega+\xi_k
\end{pmatrix}\,,
\end{equation}
where $\xi_k \equiv v_F(k-k_F)$. Due to our assumption that the sole consequence of the interaction is the Cooper pairing, 
the inverse of the 
interacting fermion propagator then takes the following form in the Nambu basis:
\begin{equation}
G^{-1}({\bf k},\omega) = \begin{pmatrix}
i \omega-\xi_k & \Delta^{*}({\bf k},\omega) \\
\Delta({\bf k},\omega) & i \omega +\xi_k
\end{pmatrix}\,.
\end{equation}
Then,
\begin{equation}
\label{eq:G}
G({\bf k},\omega) =- \frac{1}{\omega^2+\xi_k^2+ |\Delta({\bf k},\omega)|^2} \begin{pmatrix}
i \omega +\xi_k & - \Delta^{*}({\bf k},\omega) \\
- \Delta({\bf k},\omega) & i \omega-\xi_k
\end{pmatrix}\,.
\end{equation}
 
We can see now that the gap equation can be derived by inserting the above propagators into the Dyson equation.
On the Fermi surface, the gap equation comes out to be 
\begin{equation}
\Delta(\hat{\bf k},\omega) = -\int \frac{d^3 q \, d\nu}{\,(2 \pi)^4} \, V_{\text{eff}}({\bf \hat{k}}-{\bf \hat{q}},\omega-\nu)\frac{\Delta(\hat{\bf q},\nu)}{\nu^2+\xi_q^2+ |\Delta(\hat{\bf q},\nu)|^2},
\end{equation}
where
\begin{equation}
V_{\text{eff}}({\bf \hat{k}}-{\bf \hat{q}},\omega-\omega')=  \frac{( g v_F )^2}{2} \frac{1+ \cos\theta} {2 k_F^2(1-\cos \theta)+ \frac{\gamma\,|\omega-\omega'|/v_{F}}{\sqrt{2} k_F \sqrt{1 -\cos \theta}} }.
\label{EQ:effV2}
\end{equation}
$V_{\text{eff}}({\bf k},\omega; {\bf q},\nu)$ is evaluated on the Fermi surface with $\theta$ being the scattering angle: $\cos\theta = {\bf \hat{k}}\cdot{\bf \hat{q}}$. Our gap equation does not take into account the fermion wave function renormalization factor $Z$;  it has also been  dropped in  Refs.~ \onlinecite{Son:1999,Bailin:1984,Schafer:2006,Bonesteel:1999}. The infrared anomaly on the Fermi surface $\propto (\alpha/3\pi) \ln \left(\Lambda/|\omega|\right)$, shown previously,~\cite{Chakravarty:1995} is cutoff at the scale of the pairing gap. Thus, setting $Z(\omega)\approx 1$ should not be qualitatively incorrect.
 
\begin{figure}[htbp]
\begin{center}
\includegraphics[width=\linewidth]{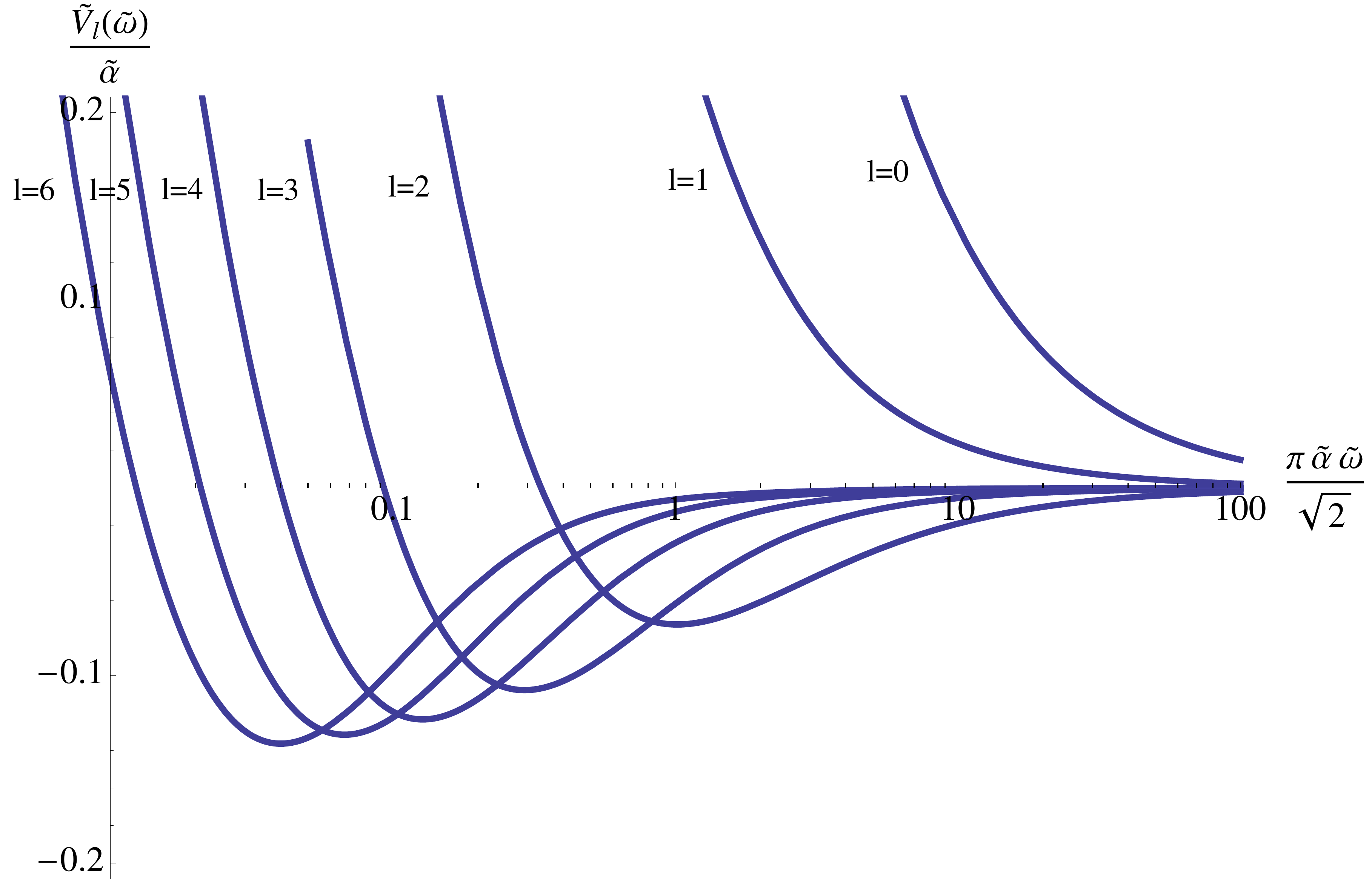}
\caption{Angular momentum decomposition of the interaction $\tilde{V}_\ell (\tilde{\omega})/\tilde{\alpha}$ versus $\pi\tilde{\alpha}\tilde{\omega}/\sqrt{2}$ for $\ell =0,1,2,3, 4,5,6.$ Note that the $x$-axis is logarithmic in $\tilde{\omega}$.}
\label{fig:vell}
\end{center}
\end{figure}

Next, we derive the self-consistency condition only with respect to the frequency by integrating out the intermediate momentum ${\bf q}$ and decompose the gap into all angular momentum channels $\ell$'s. We can transform the integral by introducing the density of states at the Fermi energy, $\int d^3 q/(2\pi)^3 \to N(0)\int d\xi_{q}\int d(\cos\theta)/2$ ($N(0) = k_F^2/\pi^2 v_F$). We obtain at $T=0$
\begin{widetext}
\begin{align}
\Delta(\hat{k},\omega) = -\alpha \int \frac{d\nu}{2\pi} \int \frac{d(\cos\theta)}{2} \frac{1+\cos\theta}{(1-\cos\theta)+\frac{\alpha}{4\sqrt{2}} \frac{|\omega-\nu|}{E_F \sqrt{1-\cos\theta}}}
\frac{\Delta(\cos\theta, \nu)}{\sqrt{\omega'^2 + |\Delta(\cos\theta, \nu)|^2}}.
\label{EQ:gapAng}
\end{align}
\end{widetext}
Being a nonlinear integral equation involving the gap, we cannot strictly speaking decouple the various angular momentum channels. However, we shall assume approximate decoupling. Such decoupled gap equations for different channels can be considered as local minima of the free energy. Therefore the self-consistency condition for the pairing gap for each angular momentum $\ell$ is, using the Fermi energy, $E_F$, as the unit energy,
\begin{equation}
\tilde{\Delta}_\ell(\tilde{\omega}) = -\int d\tilde{\nu} \tilde{V}_\ell (\tilde{\omega}-\tilde{\nu}) \frac{\tilde{\Delta}_\ell(\tilde{\nu})}{\sqrt{\tilde{\nu}^2 + |\tilde{\Delta}_\ell(\tilde{\nu})|^2}},
\label{EQ:gapAng2}
\end{equation}
where
\begin{equation}
\tilde{V}_\ell (\tilde{\omega}) = \tilde{\alpha}\int d(\cos\theta) \frac{P_\ell(\cos\theta)(1+\cos\theta)}{(1-\cos\theta)+ \frac{\pi\tilde{\alpha}}{\sqrt{2}}\frac{|\tilde{\omega}|}{\sqrt{1-\cos\theta}}},
\label{EQ:DimLess}
\end{equation}
with $\tilde{\omega} = \omega/E_F, \tilde{\nu}= \nu/E_F, \tilde{\Delta}_\ell =\Delta_\ell/E_F$ and $\tilde{\alpha} = \alpha/4\pi$. We see here that $\tilde{\alpha}$ can be treated as the dimensionless effective interaction strength. We plot $V_\ell(\omega)$ in Fig.\ref{fig:vell}. 
 
It does not require us to solve the gap equation to conclude that not only $s$-wave but also $p$-wave pairing is impossible from our Dyson formalism. As with the conventional BCS self-consistency condition, for an interaction $\tilde{V}_\ell (\tilde{\omega})$ that is always repulsive, our gap equation Eq.\eqref{EQ:gapAng2} does not have a nontrivial solution. In other words, a non-trivial solution requires $V_\ell (\omega)<0$ in at least some range of frequencies, which, according to Fig.\ref{fig:vell}, is satisfied for $\ell = 2,3,4,5,6$ etc., but not for $\ell = 0,1$. In general, we can derive from Eq.\eqref{EQ:DimLess} an identical logarithmically divergent repulsion
\begin{equation}
\tilde{V}_\ell (\tilde{\omega})\sim -\frac{2}{3}\tilde{\alpha}\log |\tilde{\omega}|,
\label{EQ:lowFreq}
\end{equation}
in the low frequency limit for all $\ell$ by making use of the fact that the forward scatter $\cos\theta \approx 1$ dominates for $|\tilde{\omega}|\ll 1$ (note, however, that this is an integrable singularity). On the other hand, in the high frequency limit $|\tilde{\omega}|\gg 1$, the Landau damping dominates, hence
\begin{equation}
\tilde{V}_\ell (\tilde{\omega}) \approx \left(\frac{\sqrt{2}}{\pi}\right)\frac{C_l}{|\tilde{\omega}|}
\label{EQ:highFreq}
\end{equation}
where $C_\ell = \int^1_{-1} P_\ell(\cos\theta)(1+\cos\theta)\sqrt{1-\cos\theta}$. We find that $C_\ell<0$ for all $\ell \geq 2$ and $C_0, C_1>0$. It is easy to see that $C_{\ell}$'s fall off rapidly with $\ell$.

A representative plot for $\tilde{V}_\ell (\tilde{\omega})$ is shown in Fig.~\ref{fig:vell}.   The $s$-and $p$-wave channels are repulsive at all frequencies, while channels for $\ell \geq 2$ are repulsive for low frequencies and attractive for higher frequencies.
The crossover frequency $\tilde{\omega}_0 (\ell,\alpha)$ for $\ell \geq 2$ can be fitted to
\begin{equation}
\tilde{\omega}_0 (\ell,\alpha)\approx  \left(\frac{\sqrt{2}}{\pi \tilde{\alpha}}\right)\frac{2.88}{\ell^{3.11}}.
\label{EQ:crossoverFreq}
\end{equation}
which indicates attractive interaction at higher frequencies for $\ell \geq 2$ angular momentum channels.  The crossover frequency could be smaller than the frequency scale at which the $Z$-factor vanishes for the fermions, as in a non-Fermi liquid, for which  only the one-loop result is available.~\cite{Chakravarty:1995} If this is the case, non-Fermi liquid effects must be directly  included within the Dyson formalism, perhaps restricting the validity of the higher angular momentum pairing.

That $p$-wave pairing is not possible 
is in contrast to the $d=2$ result~\cite{Galitski:2007} derived from the large-$N$ approximation. This is a robust result and does not depend on the Landau damping as the source of dynamic screening. A characteristic of the transverse gauge field, seen from Eqs.~\eqref{EQ:effV1} and \eqref{EQ:effV2}, is that the interaction is always suppressed for backscattering. This general feature of the transverse gauge field makes it more difficult to obtain attraction in the $p$-wave channel; obtaining attraction in the $p$-wave channel out of an interaction that is repulsive for all scattering angles -- which is exactly the case in our interaction in Eq.~\eqref{EQ:effV1} -- requires the repulsion to be stronger for backscattering than for forward scattering. Moreover, generating attraction in the $p$-wave channel through the higher order effects is not really possible in our case, unlike the Kohn-Luttinger problem.

We note that the physics of the possible pairing in the $\ell \geq 2$ channels is in some sense inverse of the BCS retardation effect \cite{Schrieffer:1963}. In the BCS case, the repulsion at higher frequency cannot break the Cooper pairs as long as there is any attraction at the zero frequency. By contrast, in this problem, $V_\ell$ for $\ell \geq 2$ gives us at low frequency a logarithmically divergent repulsion which needs to be overcome by the higher frequency attraction in order for Cooper pairing to occur, hence the requirement of the finite coupling constant. Increasing the coupling constant is advantageous for Cooper pairing not only because it strengthens the interaction but because, as we can see from Eq.\eqref{EQ:crossoverFreq}, the attraction occurs at lower frequency.

We expect the $d$-wave channel to give us the most robust pairing from Fig.~\ref{fig:vell} due to a number of different reasons. First, the $d$-wave channel has the strongest attraction in the high frequency limit, as the $|C_\ell |$ of Eq.\eqref{EQ:highFreq} falls off rapidly with $\ell$ for $\ell \geq 2$. Secondly, when one takes into account the logarithmic scale of the frequency axis of Fig.~\ref{fig:vell}, we can see that the attractive region gets narrower as $\ell$ increases. These two factors lead to
the frequency integral of $-\tilde{V}_\ell (\tilde{\omega})$ decreasing rapidly and monotonically with $\ell$ for $\ell \geq 2$ --
a result that is independent of the upper cutoff of the integral, even though the values of these integrals are cutoff dependent. 
Therefore, in the next section we will present first the numerical solution to the gap equation Eq.~\eqref{EQ:gapAng2} for $\ell=2$ and then consider $\ell > 2$ pairings.
 
\section{Numerical results}
 
\begin{figure}[htbp]
\begin{center}
\includegraphics[width=\linewidth]{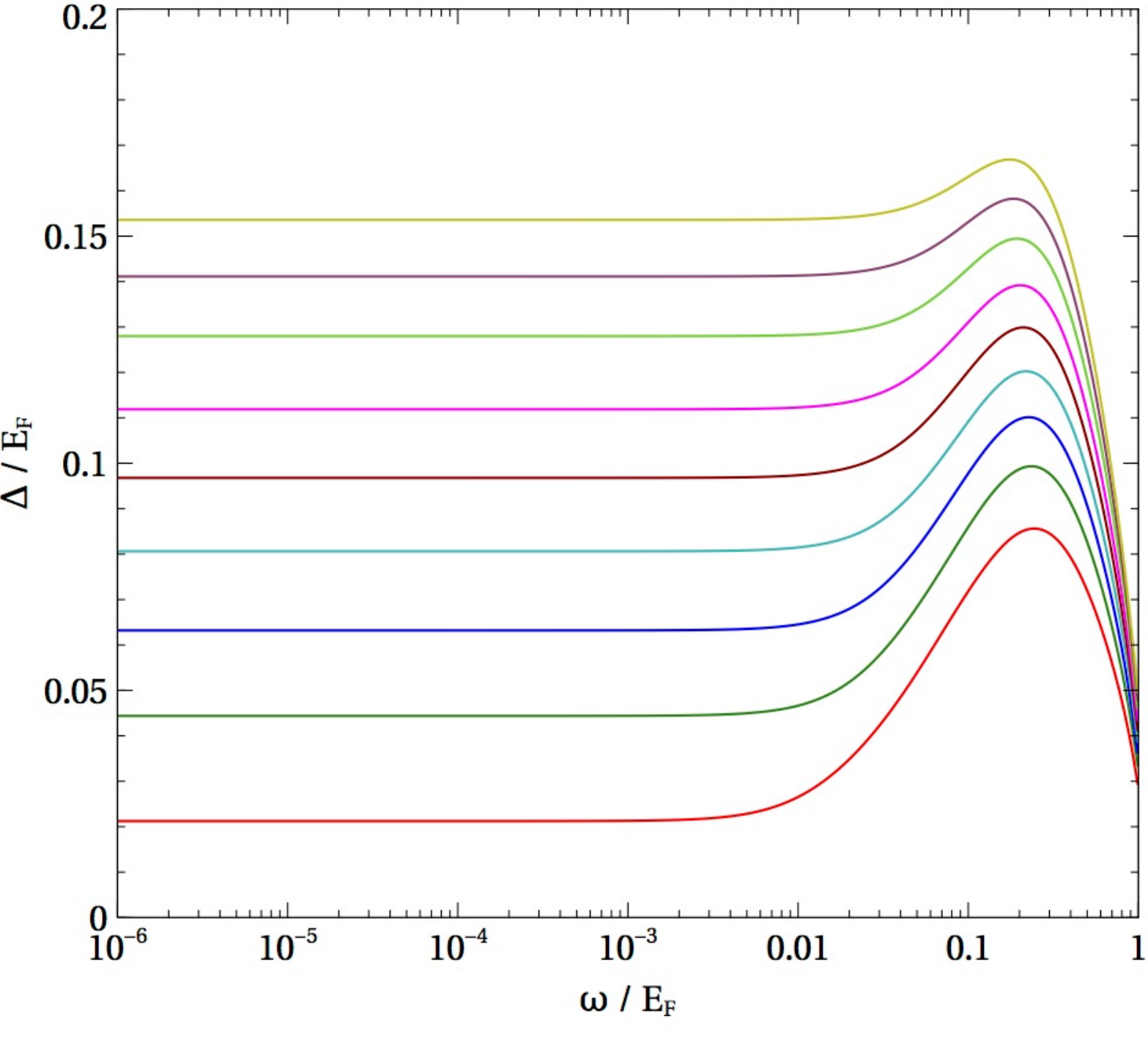}
\caption{The gap versus the imaginary frequency for the $\ell=2$ channel. From top to bottom: $\tilde{\alpha} \equiv \alpha/4\pi = 11.5, 11.0, 10.5, 10.0, 9.5, 9.0, 8.5, 8.0, 7.5$.}
\label{fig:gap-freq}
\end{center}
\end{figure}

The solution of Eq.~\eqref{EQ:gapAng2} yields a non-vanishing gap for $\alpha > \alpha_c$. To solve this equation we employed a brute force  iterative procedure with a high frequency cutoff $\omega=E_{F}$, consistent with the validity of the Landau damping. This is not always straightforward because roundoff error builds up quickly, but mostly when the gaps are small. The frequency grid also plays an important role: we have found that number of points $5\times 10^{4}$ is typically sufficient to achieve satisfactory accuracy.
The results for $\ell=2$ are shown in Fig.~\ref{fig:gap-freq}. The gap at $\omega=0$ is clearly the excitation gap, which follows from the Green function in Eq.~\eqref{eq:G} despite the Wick rotation. Obtaining gaps at real frequencies requires inverse Wick rotation.
 
Most phase transitions take place at a finite coupling which is nonuniversal. In fact, the critical coupling typically depends on the ultraviolet cutoff. A simple example to check is the large-$N$ limit for $O(N)$ nonlinear sigma model in $d=3$ where the divergence is linear in the ultraviolet cutoff.~\cite{Polyakov:1987} There are some special cases involving lower critical dimensions that are different. In one dimensional classical Ising model the critical point is exactly at $T=0$, forced to be the case because of the finite temperature fluctuations. Classical $O(3)$ sigma model at the lower critical dimension, $d=2$, behaves similarly as $T\to 0$. In these problems fluctuations at the lower critical dimension enforces the value of the critical point. 
  
\begin{figure}[htbp]
\begin{center}
\includegraphics[width=\linewidth]{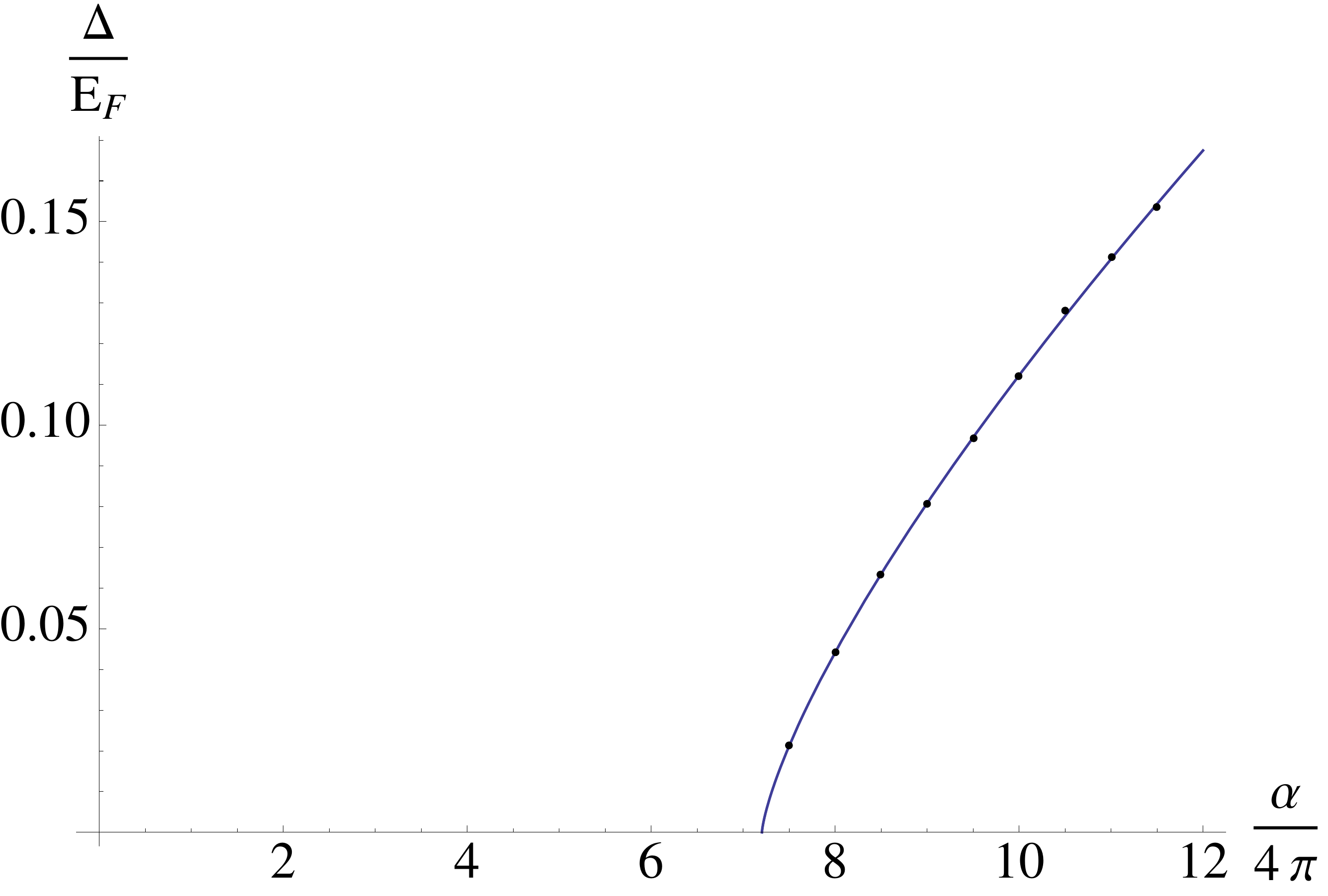}
\caption{Plot of the gap $\Delta\equiv\Delta(\omega=0)$ for the $\ell=2$ as a function of $\tilde{\alpha} \equiv \alpha/4\pi$.}
\label{fig:Delta-alpha}
\end{center}
\end{figure}

The pairing gap 
has a very strong dependence on the imaginary frequency. The superconducting phase transition, as signified by the vanishing of the $\omega=0$ gap is a continuous transition. At the edge of the superconducting phase $\alpha = \alpha_c$, the zero frequency gap 
vanish, while the finite imaginary frequency the pairing gap 
remain finite. We speculate that when analytically continued to real frequencies, there should be fluctuating gaps even when the superconductivity has disappeared. 
From Fig.~\ref{fig:Delta-alpha} we obtain the following scaling relation for the zero frequency gap in the $\ell=2$ channel:
\begin{equation}
\frac{\Delta}{E_{F}} = \begin{cases}  A_{3} \left(\frac{\alpha-\alpha_{c}}{\alpha_{c}}\right)^{0.74} & \text{if $\alpha > \alpha_{c}$,}\\
                                                                                                                                               0 & \text{if $\alpha \le \alpha_{c}$.} 
                                 \end{cases} 
\label{EQ:crit}
\end{equation}
It is unusual that the critical exponent $\approx 3/4$, which is different from the naive mean field value of $1/2$. It appears that the gap equation went beyond the mean-field formalism in the sense that our solution is non-uniform in time. As noted in the Introduction, $\alpha_{c}$ is cutoff dependent, while neither  the gap exponent nor the nature of the quantum phase transition is. For our present choice of the cutoff, $\alpha_{c}=7.2$.
 
Previous works \cite{Schafer:2006, Bonesteel:1999} have suggested that this transition is discontinuous based on various approximations to the gap equation. For example, in an insightful paper, Sch\"afer \cite{Schafer:2006} correctly pointed out that there is no superconductivity for couplings smaller than a critical coupling $\tilde{\alpha}_{c}$ from an equation, which ignores the vector character of the vertex,  the frequency dependence of the gap and assumes that the gap for all angular momentum channels are roughly the same except for $\ell=0, 1$. 
We have solved this gap equation to find that it  leads to a  jump in  the gap at $\tilde{\alpha}_{c}$ 
although the derivative at $\tilde{\alpha}_{c}+$ is singular. The jump does not imply a first order transition. There are at least two such examples: jump in the superfluid density in $2D-XY$ model\cite{Nelson:1977} and the jump in the
magnetization in the $1D$ Inverse square Ising model.~\cite{Bhattacharjee:1981} While the conclusion that a critical $\tilde{\alpha}_{c}$ is necessary for superconductivity, 
a more rigorous treatment presented here shows a gap collapsing to zero, as discussed above.

\begin{figure}[htbp]
\begin{center}
\includegraphics[width=\linewidth]{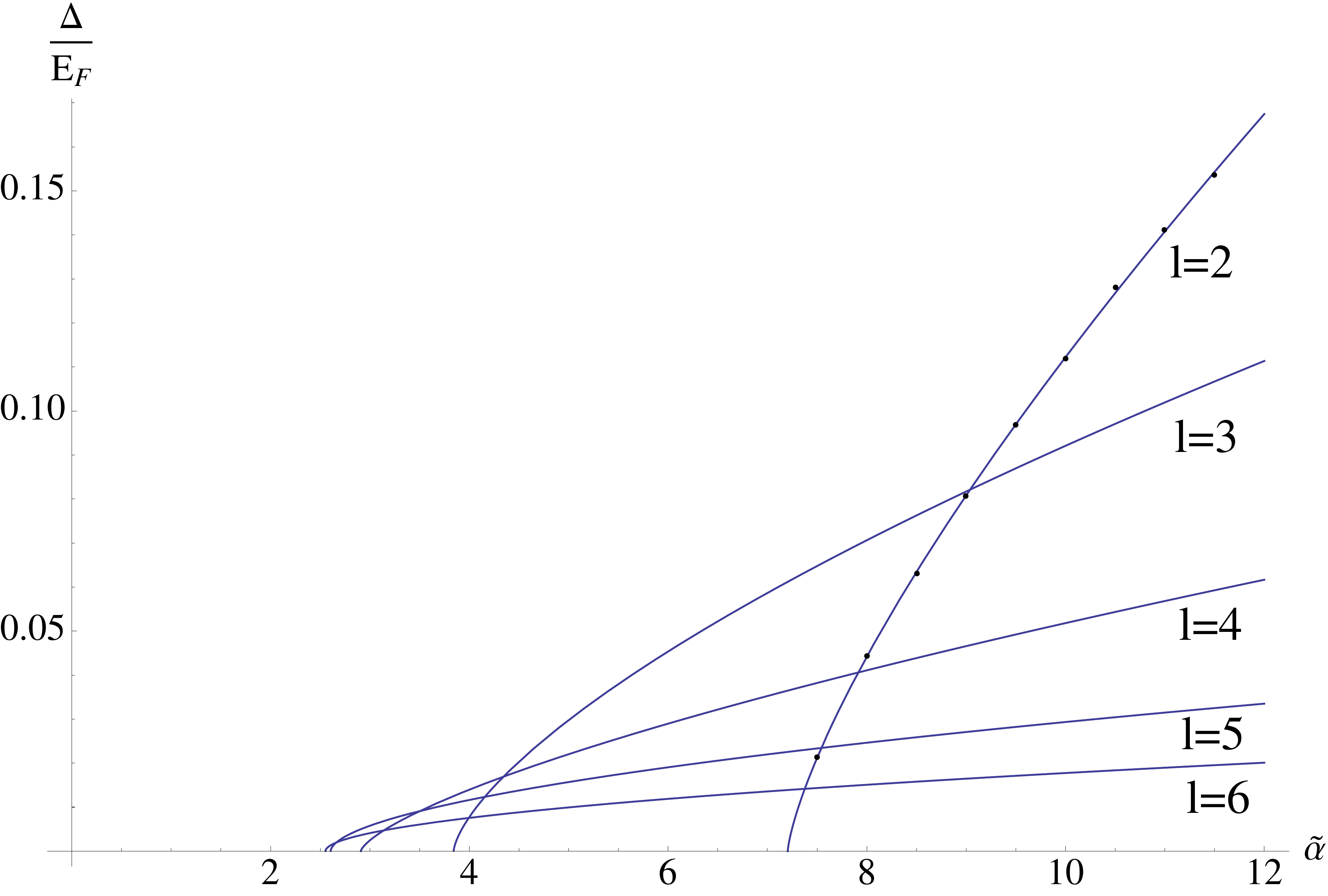}
\caption{Plot of the gap $\Delta\equiv\Delta(\omega=0)$ for the $\ell=2,3,4, 5, 6$ as a function of $\tilde{\alpha} \equiv \alpha/4\pi$.}
\label{fig:Delta-alpha-ell}
\end{center}
\end{figure}

We have further solved the gap equation for $\ell=3,4, 5, 6$ as well and find that, provided we use the same cutoff $\omega = E_F$, we obtain $\alpha_c (\ell=2)>\alpha_c(\ell=3)>\alpha_c(\ell=4)> \alpha_c (\ell=5)>\alpha_c (\ell=6)$. Fig.~\ref{fig:Delta-alpha-ell} shows plots of the zero frequency gaps for these channels as function of $\tilde{\alpha}$. It appears that that the lower bound for $\tilde{\alpha}_{c} \approx 2.5$; note that the gap equation gives us the continuous transition in $\Delta(\omega=0)$ for higher $\ell$ as well. Such a cascade of higher angular momentum gaps with an accumulation point is rather unusual. Since our Hartree-Fock equations do not include fluctuation effects, it is difficult speculate about the fate of the cascade of gaps. It is not unreasonable to expect, however, that the higher angular momentum gaps will  probably be eliminated from a variety of effects not included in our theory, but  not $\ell=2$.

\section{Conclusion}

 We have shown using the Dyson formalism that a Landau damped gauge boson in $d=3+1$ induces superconductivity with higher angular momentum pairing. No $s$-wave and $p$-wave solutions are  possible, whereas depending on the magnitude of the gauge coupling, superconducting states with $\ell \geq 2$ do occur.  Significantly, all superconducting solutions require a finite interaction strength unlike the Kohn-Luttinger instability of a Fermi liquid, which occurs for an infinitessimal repulsive interaction.  Our analysis is self-consistent in $d=3+1$, since the fermion-gauge boson coupling is marginal, and also since the superconducting gap preempts the formation of a non-Fermi liquid.  The existence of a finite critical coupling is also present in the case of finite-density quarks coupled to a non-abelian gauge field, though the starting action is quite different.   We have also carried out our analysis in the much more controversial case of $d=2+1$, and our analysis suggests again that 1) superconductivity requires a finite critical interaction, and 2) superconducting solutions have $\ell \geq 2$.    However, we have not analyzed the Chern-Simons gauge field,~\cite{Bonesteel:1999} which will be discussed elsewhere.  
 
 In this paper, we have concentrated on superconductivity mediated by transverse gauge interactions.  In this case the gauge boson is a distinct emergent field than the fermions to which it couples.  A related class of field theories involve metals in the vicinity of a Pomeranchuk instability (eg. ferromagnetism).  In this case, the boson corresponds to a fermion bilinear, and unlike the gauge field, it is generically massive except when tuned to criticality.  While superconducting instabilities in the vicinity of such quantum critical points have been studied extensively in the literature\cite{Chubukov:2005, Zaanen:2009}, many issues remain open.
 
We have not addressed the issue of the competing orders in this work. Therefore our results do not rule out Cooper pairing being pre-empted by density wave ordering in the high-$\alpha$ limit. We plan to investigate this issue through an analogous Hartree-Fock analysis.   The results of this analysis will be presented elsewhere.  
 
\begin{acknowledgments}
This work was supported by the funds from David. S. Saxon Presidential Term Chair at UCLA (IM), the Alfred P. Sloan Foundation (SR), DOE under No. AC02-76SF00515 (SR) and NSF under Grant No. DMR-1004520 (SC). We would like to thank Elihu Abrahams, Andrey Chubukov, Chetan Nayak, Joe Polchinski, and Rahul Roy for sharing their insights.
\end{acknowledgments}
 
\appendix*
\section{$(2+1)$-dimensions}

\begin{figure}[htbp]
\begin{center}
\includegraphics[width=\linewidth]{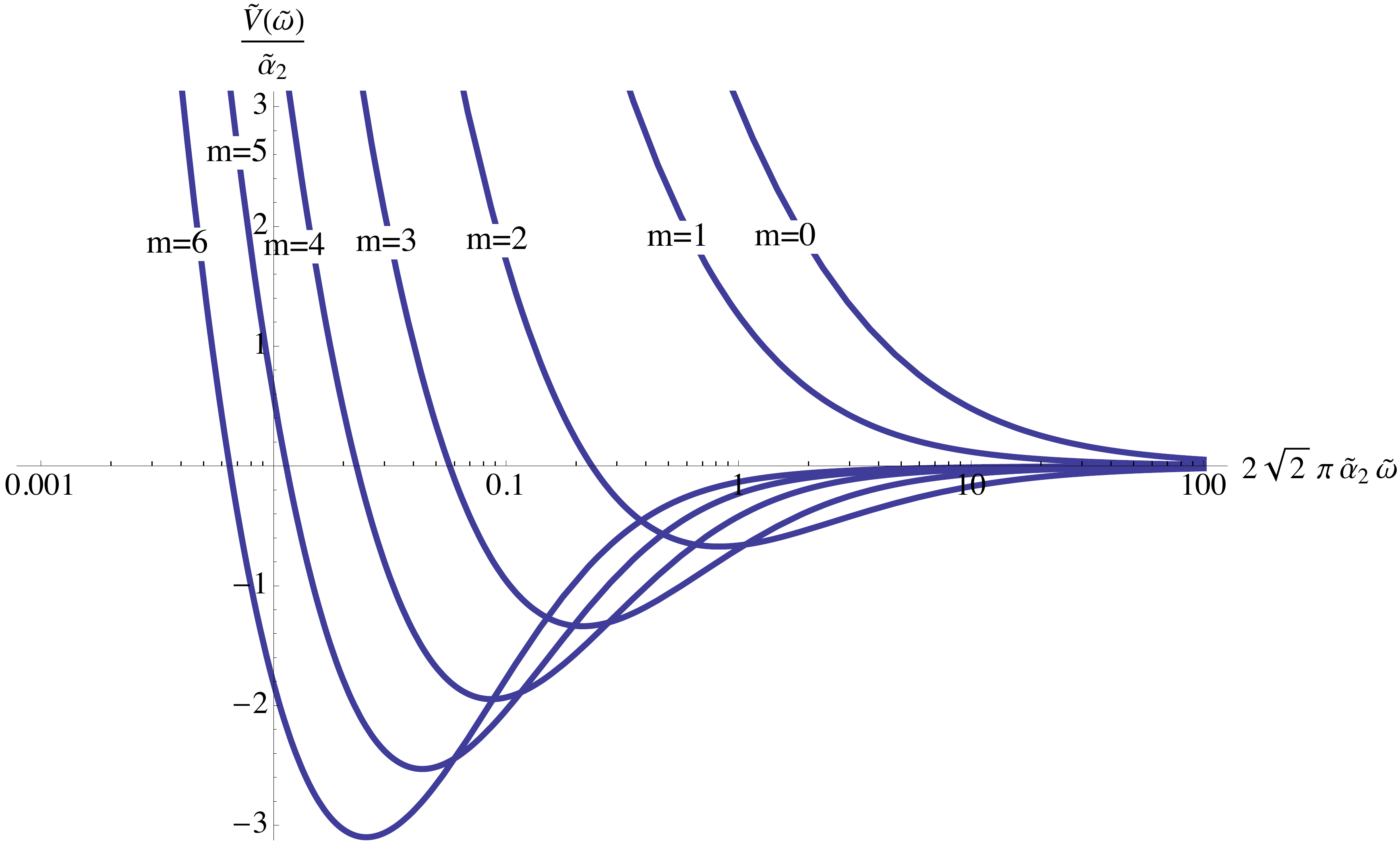}
\caption{Angular momentum decomposition of the interaction $\tilde{V}_m(\tilde{\omega})/\tilde{\alpha_{2}}$ versus $2\sqrt{2}\pi\tilde{\alpha}_{2}\tilde{\omega}$ for $m =0,1,2,3, 4,5,6$}
\label{fig:V_m}
\end{center}
\end{figure}

\begin{figure}[htbp]
\begin{center}
\includegraphics[width=\linewidth]{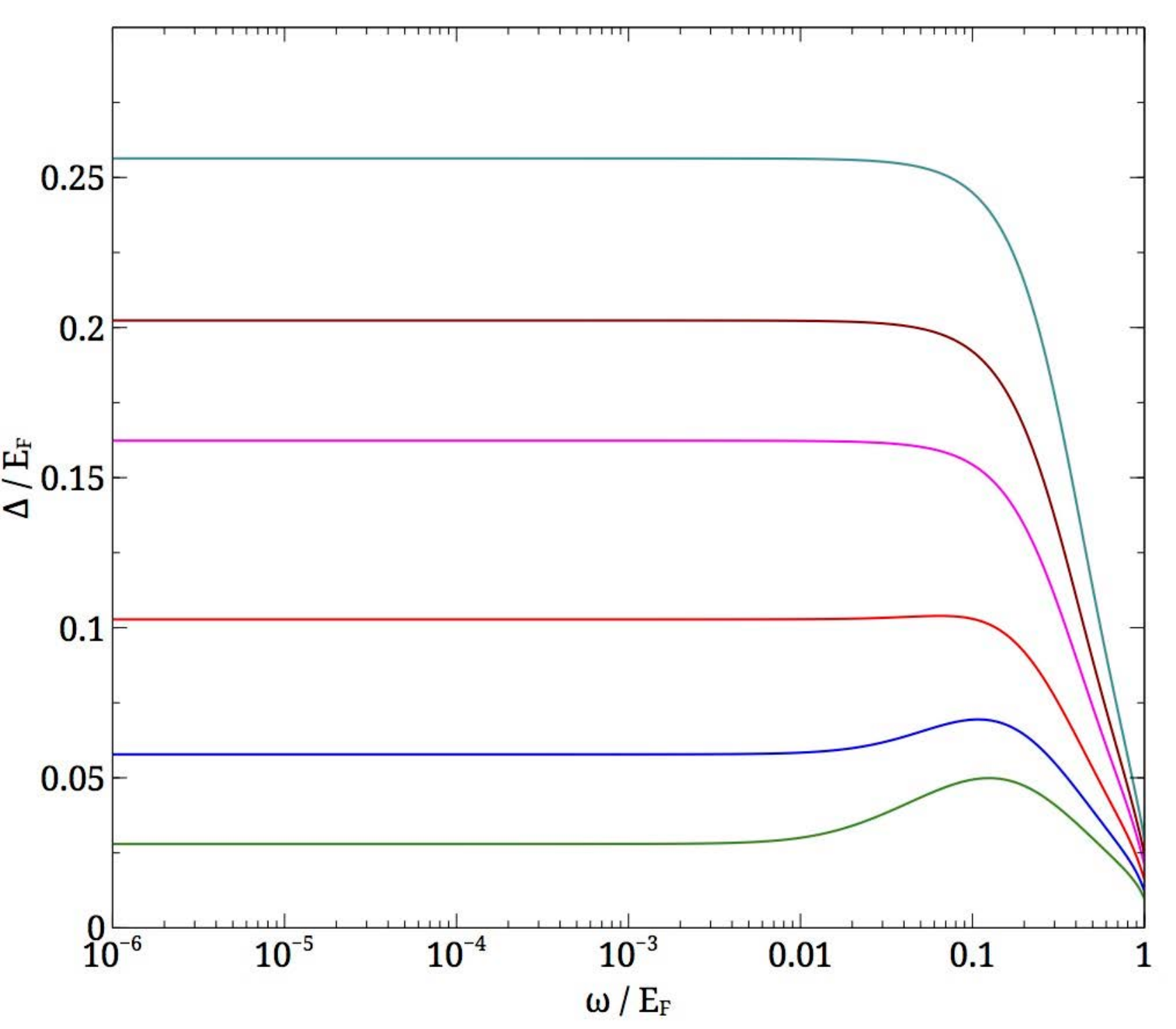}
\caption{The gap versus frequency. From top to bottom: $\tilde{\alpha}_{2} = 11.5, 7.5, 5.5, 3.5, 2.5, 2.0,$}
\label{fig:gap-freq-2}
\end{center}
\end{figure}

\begin{figure}[htbp]
\begin{center}
\includegraphics[width=\linewidth]{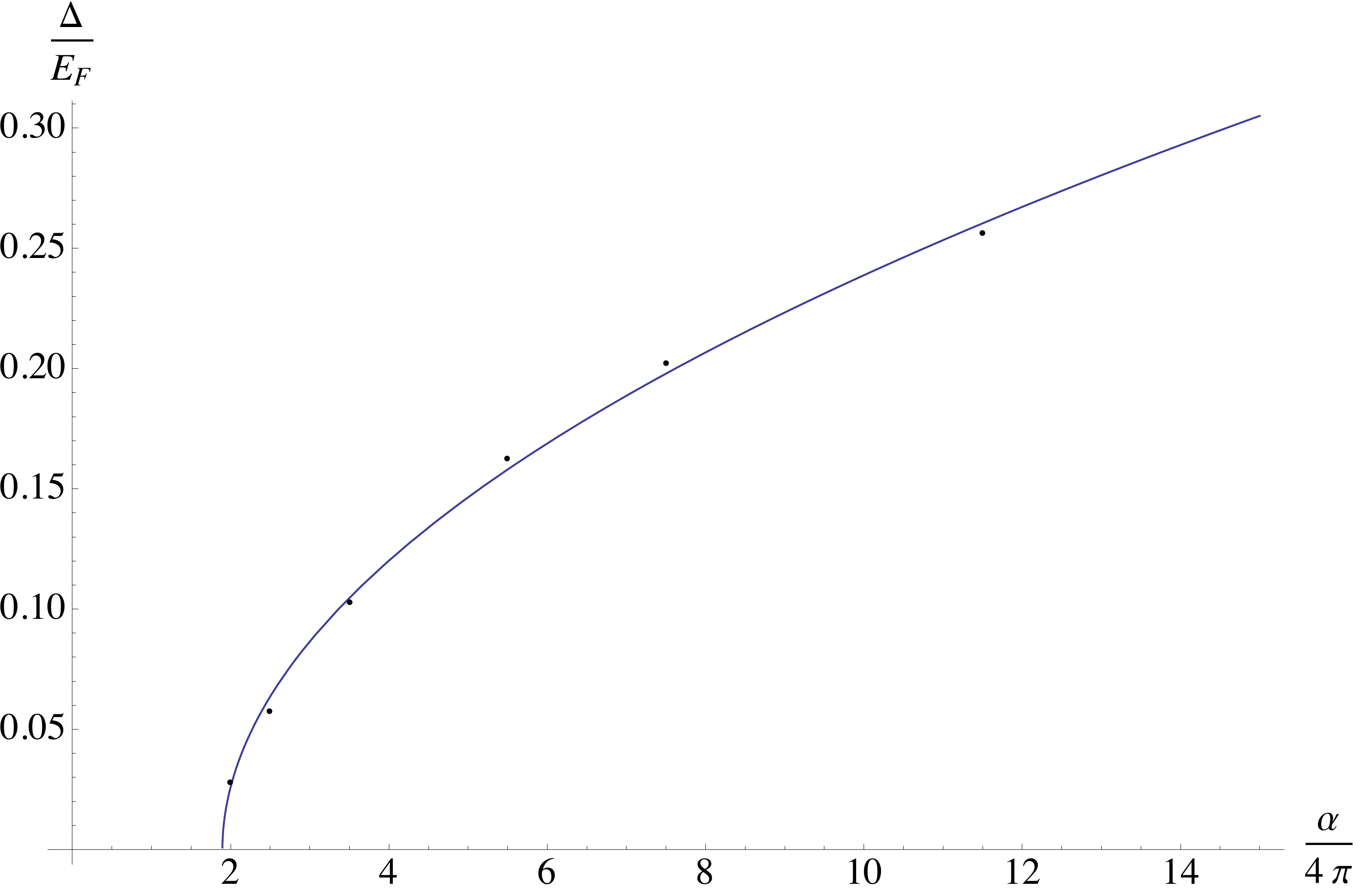}
\caption{Plot of the gap in two dimensions $\Delta\equiv\Delta(0)$ as a function of $\tilde{\alpha}$.}
\label{fig:Delta-alpha-2d}
\end{center}
\end{figure}

The non-Fermi liquid behavior in $(2+1)$  dimensions is still not well understood due to complications of the large $N$ limit~\cite{Altshuler:1994, Reply, Lee:2009,Metlitski:2010, Mross:2010}. Here we consider superconductivity from a 
Hartree-Fock approximation, with the assumption that gap might cut off the low frequency anomalies and the superconducting state may be qualitatively correct modulo fluctuation effects at the transition.  Of course, our calculation in $d=3$ is on a 
much firmer ground because $(3+1)$ is the upper critical dimension.

Because transverse gauge field canot be realized  in $(1+1)$ dimensions,  superconductivity mediated by transverse gauge bosons must have a larger  lower critical dimension. 
The gap equation for the transverse gauge boson mediated pairing interaction in $d=2$, decomposed in terms of Fourier coefficients of $\cos (m\phi)$, is
\begin{equation}
\tilde{\Delta}_m (\tilde{\omega}) = - \int d\tilde{\omega}' \, \tilde{V}_m(\tilde{\omega} -\tilde{\omega}' ) \, \frac{\tilde{\Delta}_m (\tilde{\omega}')}{\sqrt{ \tilde{\omega}'^{2} + |\tilde{\Delta}_m (\tilde{\omega}')|^2}} \,,
\end{equation}
where
\begin{equation}
\tilde{V}_m(\tilde{\omega}) = \tilde{\alpha}_{2} \int_0^{2 \pi} d \phi \frac{\cos (m\phi) (1+\cos \phi)}{ (1-\cos \phi) +\frac{\alpha_2}{\sqrt{2}} \frac{|\tilde{\omega}|}{\sqrt{1-\cos \phi}} } \,,
\end{equation}
and the dimensionless coupling constant $\tilde{\alpha}_{2}=\alpha_{2}/4\pi$:
\begin{equation}
\alpha_2 = \frac{g^2 v_F}{4 \pi k_F}.
\end{equation} 
 
In the high frequency limit $|\tilde{\omega}| \gg 1 $, the Landau damping part in the denominator dominates and we have
\begin{equation}
\tilde{V}_m(\tilde{\omega}) \approx \frac{\sqrt 2}{4 \pi} \frac{C_m}{|\tilde{\omega}|} \,,
\end{equation}
where
\begin{equation}
C_m = \int_0^{2 \pi} d \phi \cos (m\phi) (1+\cos \phi) \sqrt{1-\cos \phi},
\end{equation}
falls off, albeit less rapidly than $d=3$,  as $m$ increases, and, once again, $m=0$ and $m=1$ are repulsive. As before, the largest  gap occurs for $m=2$. Explicit solution of the gap equation in two dimensions is shown in Fig.~\ref{fig:gap-freq-2}.

The excitation gap as a function of $\tilde{\alpha}_{2}$ is shown in Fig.~\ref{fig:Delta-alpha-2d}
where the fit yields
\begin{equation}
\frac{\Delta}{E_{F}} = \begin{cases}  A_{2} \left(\frac{\tilde{\alpha}_{2}-\alpha_{c}}{\alpha_{c}}\right)^{0.51} & \text{if $\tilde{\alpha}_{2} > \alpha_{c}$,}\\ 
                                                                                                                                               0 & \text{if $\tilde{\alpha}_{2} \le \alpha_{c}$.}
                                 \end{cases}
\label{EQ:crit2d}
\end{equation}
where $\alpha_{c}=1.89$ is of course cutoff dependent and $A_{2}$ is a dimensionless constant. 

\begin{figure}[htbp]
\begin{center}
\includegraphics[width=\linewidth]{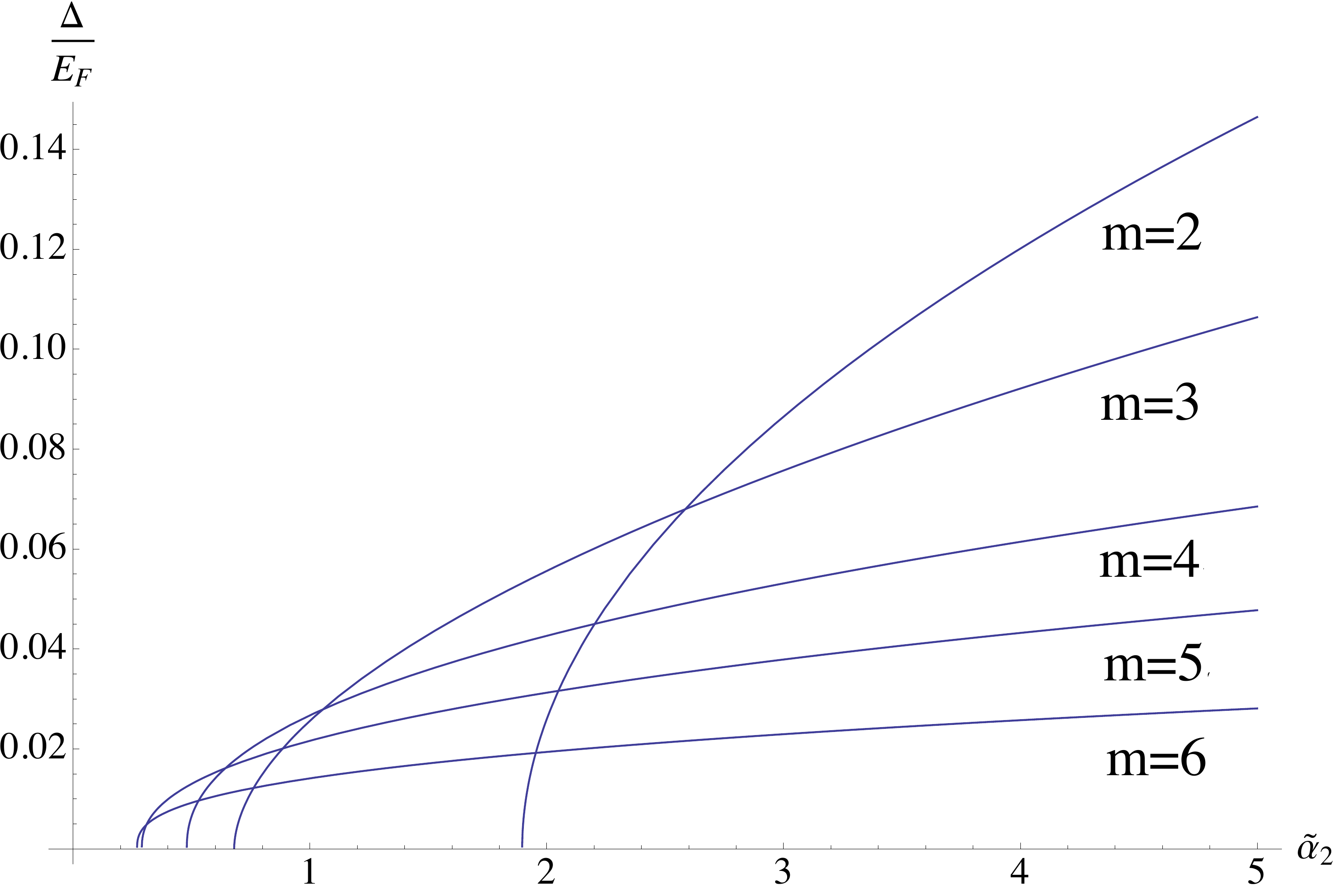}
\caption{Plot of the gap in two dimensions $\Delta\equiv\Delta(0)$ as a function of $\tilde{\alpha}$ for $m= 2, 3, 4, 5, 6$. The lower bound for $\tilde{\alpha}_{c}\approx 0.27$.}
\label{fig:Delta-alpha-2d-all}
\end{center}
\end{figure}

We have also solved the gap equation for $m=3, 4, 5, 6$ and find that the lower bound for  $\tilde{\alpha}_{c}$ is non-vanishing as for $d=3$, and the picture of cascade higher angular momentum states is identical. The results are shown in Fig.~\ref{fig:Delta-alpha-2d-all}.

\end{document}